\documentstyle[12pt]{article}
\def\ftoday{{\sl  \number\day \space\ifcase\month
\or Janvier\or F\'evrier\or Mars\or avril\or Mai
\or Juin\or Juillet\or Ao\^ut\or Septembre\or Octobre
\or Novembre \or D\'ecembre\fi
\space  \number\year}}

\makeatletter
\@addtoreset{equation}{section}
\makeatother

\newcommand{\es}{\\[3mm]}

\renewcommand{\a}{\alpha}
\renewcommand{\b}{\beta}
\newcommand{\g}{\gamma}           \newcommand{\G}{\Gamma}
\renewcommand{\d}{\delta}         \newcommand{\D}{\Delta}
\newcommand{\e}{\varepsilon}

\newcommand{\la}{\lambda}        
\newcommand{\m}{\mu}
\newcommand{\n}{\nu}
\newcommand{\om}{\omega}         \newcommand{\OM}{\Omega}
             
\newcommand{\r}{\rho}
\newcommand{\s}{\sigma}           \renewcommand{\S}{\Sigma}
\newcommand{\th}{\theta}         
\newcommand{\f}{{\phi}}

\newcommand{\z}{\zeta}
\renewcommand{\AA}{{\cal A}}
\newcommand{\BB}{{\cal B}}

\newcommand{\GG}{{\cal G}}
\newcommand{\HH}{{\cal H}}

\newcommand{\NN}{{\cal N}}

\newcommand{\PP}{{\cal P}}
\newcommand{\QQ}{{\cal Q}}

\newcommand{\UU}{{\cal U}}
\newcommand{\VV}{{\cal V}}
\newcommand{\WW}{{\cal W}}

\def\Lp{\displaystyle{\biggl(}}
\def\Rp{\displaystyle{\biggr)}}

\newcommand{\lp}{\left(}\newcommand{\rp}{\right)}
\newcommand{\lc}{\left[}\newcommand{\rc}{\right]}
\newcommand{\lac}{\left\{}\newcommand{\rac}{\right\}}

\newcommand{\complex}{{\kern .1em {\raise .47ex
\hbox {$\scriptscriptstyle |$}}
    \kern -.4em {\rm C}}}
\newcommand{\real}{{{\rm I} \kern -.19em {\rm R}}}
\newcommand{\rational}{{\kern .1em {\raise .47ex
\hbox{$\scripscriptstyle |$}}
    \kern -.35em {\rm Q}}}
\renewcommand{\natural}{{\vrule height 1.6ex width
.05em depth 0ex \kern -.35em {\rm N}}}

\newcommand{\tr}{{\rm {Tr} \,}}
\newcommand{\half}{\frac 1 2}
\newcommand{\pa}{\partial}

\newcommand{\dpad}[2]{{\displaystyle{\frac{\partial #1}{\partial #2}}}}
\newcommand{\dfud}[2]{{\displaystyle{\frac{\delta #1}{\delta #2}}}}

\newcommand{\dint}{\displaystyle{\int}}

\newcommand{\sla}{\raise.15ex\hbox{$/$}\kern -.57em}
\newcommand{\twiddle}{\lower.9ex\rlap{$\kern -.1em\scriptstyle\sim$}}

\newcommand{\equ}[1]{(\ref{#1})}
\newcommand{\eq}{\begin{equation}}
\newcommand{\eqn}[1]{\label{#1}\end{equation}}
\newcommand{\eea}{\end{eqnarray}}
\newcommand{\eqa}{\begin{eqnarray}}
\newcommand{\eqan}[1]{\label{#1}\end{eqnarray}}
\newcommand{\ba}{\begin{array}}
\newcommand{\ea}{\end{array}}
\newcommand{\eqac}{\begin{equation}\begin{array}{rcl}}
\newcommand{\eqacn}[1]{\end{array}\label{#1}\end{equation}}
\newcommand{\qq}{&\qquad &}
\renewcommand{\=}{&=&} 
\begin{document}
\begin{titlepage}
\topskip1cm
\flushright{TUW-93-22}\\[3cm]
\begin{center}{\huge\bf
Interpolating gauge fixing for Chern-Simons theory\\[2cm]}{
\Large K. Landsteiner, M. Langer,\\M. Schweda and S.P. Sorella
\\[1cm]}
Institut f\"ur Theoretische Physik\\
Technische Universit\"at Wien\\
Wiedner Hauptstra{\ss}e 8-10\\
1040 Wien (Austria)\\[2cm]
\end{center}
\begin{abstract}
Chern-Simons theory is analyzed with a gauge-fixing which allows to discuss
the Landau gauge and the light-cone gauge at the same time.
\\[15mm]
\end{abstract}
September 1993
\end{titlepage}
%
%
\section{Introduction}
The Chern-Simons theory in three space-time dimensions \cite{witten}
has been the
object of continuous investigations during the last years.
In particular, its topological nature has led to the beautiful and
powerful relation between topological invariants of three dimensional
manifolds and the vacuum expectation value of
Wilson lines \cite{witten,guad,labast}.

The Chern-Simons theory has been extensively studied also as a pure
three-dimensional gauge field model. It turns out to be an
example of a fully ultraviolet finite theory \cite{dlps,bc}.
This is due to the
existence, besides the usual gauge invariance, of an additional
symmetry, usually called topological vector-supersymmetry, whose
generators carry a Lorentz index. This supersymmetry, originally found
in the covariant Landau gauge \cite{brt,dgs}, has been recently observed
also in the noncovariant algebraic gauges \cite{blsps,kumm}
and imposes quite strong constraints
on the theory. Indeed, as it has been shown in the
case of the axial gauge \cite{blssep}, it allows to compute
exactely all the
Green functions without the need of an action principle and of the usual
Feynman graphs expansion derived from it.

Let us mention also that, actually, the topological supersymmetry seems
to be a common feature of a very large set of topological field
theories \cite{susy}
including the bosonic string \cite{corde}.
Moreover, as proven in \cite{silvio}, it provides
an elegant and simple way for solving the descent equations associated to the
integrated BRST cohomology, yielding then an algebraic characterization of
anomalies and invariant counterterms.

The aim of this letter is to examine the symmetry content of the
Chern-Simons theory in a more general gauge known as the
{\it interpolating gauge}. This gauge, introduced by \cite{pps} in the
case of four-dimensional Yang-Mills theories, interpolates between the
Landau gauge and the light-cone gauge with the Leibbrandt-Mandelstam
prescription \cite{leibb}. This is done by means of the introduction of
a gauge
parameter $\z$ running from $0$ to $\infty$, the two values correspond
respectively to the Landau gauge $(\z=0)$ and to the the light-cone
gauge $(\z=\infty)$.

The main advantage of the use of the interpolating gauge in four
dimensional gauge theories relies on the
possibility of regularizing the nonlocal divergences which arise when
pure noncovariant gauges are adopted \cite{leibb}. Indeed, as discussed in
details in \cite{pps}, for any
finite value of the gauge parameter $\z$ the ultraviolet behaviour of
the propagators turns out to be compatible with the requirement of
the power-counting theorem. This implies that only local
counterterms are needed in order to renormalize the theory.
The limit $\z \rightarrow \infty$ is, of course, singular and one recovers
the usual nonlocal divergences. However, being $\z$ a gauge
parameter, it follows that the matrix element of gauge invariant
operators does not depend on $\z$, i.e. the limit $\z \rightarrow \infty$
exists. One sees then that the use of the interpolating gauge allows
to prove in an algebraic way the important result that in a four-dimensional
gauge theory the nonlocal divergences drop out in the matrix element
of physical quantities.

Let us turn now to the case of the three-dimensional Chern-Simons
theory. Anticipating the conclusions, we may say that the use
of the interpolating gauge gives quite interesting results also
in the case of the topological field  theories.
In particular, as we shall see in the following, the topological
supersymmetry holds also in the interpolating gauge.
We will be able then to prove that, as in the case of a pure Landau gauge
\cite{dlps,bc}, the model is ultraviolet finite.
Therefore, contrary to the four-dimensional case, the limit
$\z \rightarrow \infty$ is not singular.

This feature has an important consequence: it implies that the transition
between a covariant Landau gauge and a noncovariant algebraic gauge
can be done in a soft way, i.e. in the case of topological field theories
the use of the interpolating gauge allows to reach smoothly an algebraic
noncovariant gauge starting from a covariant one. In addition it
suggests also that the matrix
element of Wilson loops, being a gauge invariant quantity, does not depend
on the gauge parameter $\z$ and then its value remains unchanged when one
moves from $\z=0$ to $\z=\infty$. This result has been confirmed
by \cite{fk} with a different technique.

The work is organized as follows. In Sect.2 we present the quantization
of the model in the interpolating gauge, in Sect.3 we establish the
topological vector supersymmetry and, finally, in Sect.4 we discuss the
ultraviolet finiteness of the model.
%
%
\section{The classical action in the interpolating gauge}
The classical gauge-fixed action we start with is given by
\eq
   S=S_{inv}+S_{gf} \ ,
\eqn{CSges}
where $S_{inv}$ and $S_{gf}$ are respectively the gauge invariant
Chern-Simons action and the interpolating gauge-fixing term \cite{pps}.
They read
\eq
   S_{inv}=-\frac{1}{2k} \dint\,d^3x\, \tr\e^{\m\n\r}\Lp A_\m\pa_\n A_\r
    +\frac{2}{3} A_\m A_\n A_\r \Rp ,
\eqn{CSinv}
and
\eqa
  S_{gf}=s\dint\,d^3x\, \tr \Lp \bar{K}\lp\z(n^\ast\pa)D-(\pa A)\rp
     +\bar{c}\lp D + nA\rp \Rp\, ,
\eqan{CSfix}
where $s$ denotes the BRST-transformations
\eqac
  sA_\m=-D_\m c\,,\qq sc=c^2\,,\es
  s\bar{c} =b\,,\qq sb=0\,,\es
  s\bar{K}=K\,,\qq sK=0\,,\es
  sD=\hat{D}\,,\qq s\hat{D}=0\, ,\es
\eqacn{BRS-Tr}
with
\eq
   D_\m c = \pa_\m c + [A_\m,c]    \ .
\eqn{covderiv}
All the fields $\phi$ are Lie algebra-valued $\phi = \phi^a T^a$,
$T^a$ being the generators of a simple gauge-group.
We use a Minkowskian metric
$\eta_{\m\n}={\rm diag.}(+,-,-)$ and we assume that
$n^\m$ is a light-like vector $(|\vec{n}|,\vec{n})$ with $n^{*\m}$
the dual vector $(|\vec{n}|,-\vec{n})$.
The gauge-fixing \equ{CSfix} interpolates between the Landau-gauge
($\zeta = 0$) and the light-cone gauge ($\zeta \rightarrow \infty$) as
one can see from the expressions of the propagators. The latters
are computed to be

\begin{equation}
\lp \begin{array}{cccc}
\e_{\a\m\n}\frac{k}{\pa_\z^2}\lp \pa^\a+ n^\a \z(n^\ast \pa)\rp &
-\frac{\z(n^\ast\pa)}{\pa^2_\z }\pa_\n &
\e_{\a\b\n}\frac{n^\b\pa^\a}{\pa^2_\z} &
\frac{\pa_\n}{\pa^2_\z}\es
-\frac{\z({n^\ast}\pa)}{\pa^2_\z }\pa_\m &
0 &
-\frac{\pa^2}{\pa_\z^2} &
0 \es
\e_{\a\b\m}\frac{n^\b\pa^\a}{\pa^2_\z} &
-\frac{\pa^2}{\pa_\z^2} &
0 &
\frac{(n\pa)}{\pa_\z^2}\es
\frac{\pa_\m}{\pa^2_\z} &
0 &
\frac{(n\pa)}{\pa_\z^2} &
0\es
\end{array}\rp\d(x-y)\d^{ab}\,,\\ \label{fprop}\end{equation}

for the $ (A_\m\ ,b\ ,D\ ,K)$ sector and

\begin{equation}
\lp \begin{array}{cccc}
0 &
-\frac{\z(n^\ast\pa)}{\pa_\z^2} &
0 &
\frac{1}{\pa_\z^2}\es
-\frac{\z(n^\ast\pa)}{\pa_\z^2} &
0 &
\frac{\pa^2}{\pa_\z^2} &
0 \es
0 &
-\frac{\pa^2}{\pa_\z^2} &
0 &
\frac{(n\pa)}{\pa_\z^2}\es
-\frac{1}{\pa_\z^2} &
0 &
\frac{(n\pa)}{\pa_\z^2} &
0 \es
\end{array}\rp\d(x-y)\d^{ab}.\\ \label{gprop}\end{equation}
for the ghost sector $(c\ ,\bar{c}\ ,\hat{D}\ ,\bar{K})$
with
\eq
  \pa^2_\z = \pa^2 + \z (n^*\pa)(n\pa) \ .
\eqn{partial-z}
It is easy now to recover the propagators of the
Landau-gauge for $\z = 0$ and of the light-cone gauge with the
Leibbrandt-Mandelstam prescription for
$\z \rightarrow \infty$ .\\
Notice that, as already discussed in \cite{pps}, for any finite value
of $\z$ all the relevant propagators are compatible
with the requirement of the power-counting theorem.
Therefore we are allowed to study the ultraviolet behaviour of the model
by making use of the Quantum Action Principle \cite{qap} which, in turn,
guarantees that all possible counterterms and anomalies are
local functionals of the fields.
Let us remark also that the parameter $\z$, due to the fact that
it appears only in the pure gauge-fixing part of the action \equ{CSges},
is a gauge parameter.
%
%
\section{Symmetries}
It is well known that the Chern-Simons theory, when quantized in
the pure Landau gauge \cite{brt,dgs} or in a noncovariant algebraic
gauge \cite{blsps,kumm}, possesses a
vector-supersymmetry which turns out to be very powerfull for
studying its perturbative behaviour.
Let us show that this supersymmetric
structure is present also in the interpolating gauge \equ{CSfix}.
In order to derive it let us consider, following \cite{ms}, the
energy-momemtum tensor $T_{\mu\nu}$ which, due to the topological
character of
the action \equ{CSges}, turns out to be a BRST-variation
\eq
    T_{\mu\nu} = s \Lambda_{\mu\nu} \ ,
\eqn{energy-momentum}
with
\eqac
\Lambda_{\mu\nu} \=
    +\eta_{\m\n}\Lp\z\bar{K}(n^\ast \pa) D+
   (A\pa) \bar{K}  + \bar{c}(n A) +\bar{c} D\Rp \es
 &&   -\lp\z\bar{K}n_\m^\ast \pa_\n D+
    \pa_\m \bar{K} A_\n + \bar{c}n_\m A_\n
   + \m\leftrightarrow \n\rp\ \ . \es
\eqacn{Lambdamn}
In particular the divergence of $\Lambda_{\mu\nu}$ reads
\eqac
\pa^\m\Lambda_{\m\n} \=\tr\Lp \pa_\n{\bar c}\dfud{S}{b} + \pa_\n{\bar K}
\dfud{S}{K}+\pa_\n D\dfud{S}{\hat{D}}+ \e_{\m\n\r}(\pa^\m{\bar K}+
n^\m{\bar c})\dfud{S}{A_\r}- A_\n\dfud{S}{c}\Rp \es
&&+s\tr \Lp\e_{\m\n\r} n^\m\bar{c}\pa^\r \bar{K}\Rp +
      \mbox{{tot. deriv.}}\es
\eqacn{paLambda}
so that, integrating on space-time one gets the equation
\eq\ba{rl}
   \dint\,d^3x \,\tr &\!\! \Lp \pa_\n{\bar c}\dfud{S}{b} + \pa_\n{\bar K}
    \dfud{S}{K}+\pa_\n D\dfud{S}{\hat{D}} +
    \e_{\m\n\r}(\pa^\m{\bar K} + n^\m{\bar c})\dfud{S}{A_\r}
   -  A_\n\dfud{S}{c}\Rp = \\
 &\!\!
   - s\dint\,d^3x\,\tr \Lp\e_{\m\n\r} n^\m\bar{c}\pa^\r \bar{K}\Rp  \ .
\ea\eqn{integr-identity}
This identity, due to the fact that the quadratic breaking in the
right hand side is a BRST variation, can be converted into an
exact symmetry of the classical action.
This is easily done by introducing a BRST doublet of external sources
$(\la^\nu, {\bar \la}^\nu)$  which are singlet with respect to the gauge
group
\eq
    s\la^\m ={\bar \la}^\m {\ }{\ }{\ }\ ,{\ }{\ }{\ }
    s{\bar \la}^\m =0 \ .
\eqn{newsBRS}
Adding now to the classical action $S$ of \equ{CSges} the external
coupling
\eq
   S^{(\la)} = \tr\dint\,d^3x\,
      s\lp \e_{\m\n\r}\la^\m n^\n \bar{c} \pa^\r \bar{K}\rp \, .
\eqn{Slambda}
one easily checks that the modified action
\eq
   {\hat S} = S + S^{(\la)} \ ,
\eqn{actmod}
is left invariant by the following transformations
\eqac
\d_\m A_\n=\e_{\m\n\r}(\pa^\r\bar{K}+n^\r \bar{c} ),\qq
\d_\m b=\pa_\m \bar{c},\es
\d_\m K=\pa_\m \bar{K}, \qq \d_\m\hat{D} =\pa_\m D,\es
\d_\m {\bar \la}^\n =\pa_\m \la^\n, \qq \d_\m c=-A_\m, \es
\d_\m \la^\n =\d_\m^{\  \n}, \qq \d_\m \bar{c}=0 ,\es
\d_\m D=0, \qq \d_\m \bar{K}=0 ,\es
\eqacn{newsw}
and
\eqac
\d_\m {\hat S} = 0\,,
\eqacn{onsusy}
The operator $\d_\mu$ gives rise, together with
the BRST operator, to the following on-shell supersymmetric structure
\eq\ba{lcl}
   \lac s,\d_\m\rac \f = \pa_\m \f {\ }+{\rm eq.}{\ }{\rm of}
                                        {\ }{\rm motion}\ , \\
   \lac \d_\m , \d_\n \rac = s^2 = 0 \ ,
\ea\eqn{acswm}
where $\f$ collects all the fields.

Finally, coupling the non-linear BRST-variations of
$A_\m$ and $c$ to the external sources $\g^\m$ and $\s$,
\eqac \S = {\hat S} + \tr\dint\,d^3x\,
\Lp  -\g^\m \lp\pa_\m c+ \lc A_\m ,c\rc \rp +\s c^2\Rp \, .\es
\eqacn{acttot}
one gets the classical Slavnov identity
\eqac
\BB(\S) \= \tr \dint\,d^3x\, \Lp\dfud{\S}{\g^\m}\dfud{\S}{A_\m}+
\dfud{\S}{\s}\dfud{\S}{c}
+b\dfud{\S}{\bar{c}}+
\hat{D}\dfud{\S}{D}+K\dfud{\S}{\bar{K}}
 + {\bar \la}^\m\dfud{\S}{\la^\m} \Rp =0, \es
\eqacn{slatay}
and the supersymmetric Ward identity
\eq
\WW_\m \S = \D^{cl}_\m
\eqn{susy}
with
\eq\ba{rl}
\WW_\m = \tr\dint\,d^3x\,\Lp &\!\!
\e_{\m\n\r}(\pa^\r\bar{K}+n^\r \bar{c} - \g^\r) \dfud{}{A_\n}
- A_\m \dfud{}{c} +
\pa_\m \bar{c} \dfud{}{b} + \pa_\m \bar{K} \dfud{}{K} \\
 &\!\! +\pa_\m D \dfud{}{\hat{D}}
       + \pa_\m \la^\n \dfud{}{{\bar \la}^\n} +
       \dfud{}{\la^\m} -\s  \dfud{}{\g^\m}\Rp \ ,
\ea\eqn{susy-operator}
and
\eq
 \D^{cl}_\m = \tr\dint\,d^3x\,
\Lp \s\pa_\m c - \g^\n\pa_\m A_\n
 - \e_{\a\m\b}\g^\a(\pa^\b K  + n^\b b)\Rp \ .
\eqn{clbrk}
Notice that the breaking $\D^{cl}_\m$, being linear in the quantum
fields, is a classical breaking.

Introducing now the nilpotent linearized Slavnov operator
\eq\ba{rl}
  \BB_{\S} = \tr\dint\,d^3x\, \Lp &\!\!
  \dfud{\S}{\g^\m}\dfud{}{A_\m}+
  \dfud{\S}{A_\m}\dfud{}{\g^\m}+
  \dfud{\S}{\s}\dfud{}{c}+
  \dfud{\S}{c}\dfud{}{\s}   \\
  &\!\! +b\dfud{}{\bar{c}}+
  \hat{D}\dfud{}{D}+K\dfud{}{\bar{K}}+
  {\bar \la}^\m\dfud{}{\la^\m}\Rp  \ ,
\ea\eqn{lslatay}
we get the off-shell Wess-Zumino type algebra
\eqac
\BB_{\S}\BB_{\S} = 0\,,\es
\lac\WW_\m , \WW_\n \rac = 0\,,\es
\lac\WW_\m , \BB_{\S} \rac = \PP_\m\,,\es
\eqacn{offsusy}
where $\PP_\m$ is the translation Ward operator
\eq
  \PP_\m = \sum_{all{\ }fields{\ }\f} \tr\dint\,d^3x\,
                     \pa_\m \f \dfud{ }{\f} \ .
\eqn{translations}
We remark that the supersymmetric Ward operator $\WW_\m$ doesn't depend
explicitely on the interpolating gauge parameter $\z$, implying that the
supersymmetric structure \equ{offsusy} remains unchanged when one performs
the limits $\z=0$ and $\z=\infty$. This is in agreement with the results
obtained in \cite{dgs,blsps}.

Besides the Slavnov identity \equ{slatay} and the supersymmetric
Ward identity \equ{susy}, the classical action \equ{acttot} turns
out to be constrained also by: \es
(i) four gauge conditions
\eqac \qq \dfud{\S}{b} = D+(nA)+\e_{\m\n\r}n^\m\pa^\n\bar{K} \la^\r,\es
      \qq \dfud{\S}{D}=-\z (n^\ast \pa)K +b,\es
      \qq \dfud{\S}{\hat{D}} =-\z (n^\ast\pa) \bar{K} +\bar{c},\es
    \qq \dfud{\S}{K}= \z (n^\ast\pa)D- (\pa A) -\e_{\m\n\r}\pa^\m(\la^\n n^\r
\bar{c}),\es
\eqacn{gaugecon}
(ii) two anti-ghost equations
\eqac \qq \dfud{\S}{\bar{K}}-\pa^\m\dfud{\S}{\g^\m}= -\z(n^\ast\pa)
\hat{D} + \e_{\m\n\r} \pa^\m({\bar \lambda}^\n n^\r \bar{c}
                             - \la^\n n^\r b),\es
   \qq \dfud{\S}{\bar{c}} +n^\m \dfud{\S}{\g^\m}=-\hat{D} + \e_{\m\n\r}
n^\m  ({\bar \lambda}^\r \pa^\n \bar{K}- \la^\r \pa^\n K ), \es
\eqacn{antighost}
(iii) the ghost equation, usually valid in the Landau gauge \cite{ghost},
in this
case extends to
\eqac \GG \S\=  \dint\,d^3x\, \Lp \dfud{\S}{c} + \lc\bar{K},\dfud{\S}{K}\rc
 + \lc \bar{c},
\dfud{\S}{b}\rc + \lc D,\dfud{\S}{\hat{D}}\rc \Rp \es\=
 \dint\,d^3x\,\Lp \lc c,\s\rc -
\lc A_\m,\g^\m\rc\Rp, \es
\eqacn{ghost}
(iv) two transversality conditions
\eqac  \qq \UU\S=\dint\,d^3x\, n^\m \dfud{\S}{\la^\m}=0,\es
       \qq \VV\S=\dint\,d^3x\, n^\m \dfud{\S}{{\bar \lambda}^\m}=0,
\eqacn{transcon}
which express the fact that the composite operators defined in \equ{Slambda}
are orthogonal to the vector $n^\m$, \es
(v) the rigid gauge-invariance
\eqac \HH^a\S\=  \dint\,d^3x\,\sum_{all{\ }fields{\ }\f}
       f^{abc}\f_b\dfud{\S}{\f_c}.
\eqacn{rigcon}
In addition, one has the following algebraic relations
\eqac
   \lac \GG^a , \GG^b \rac = 0 \ , {\ }{\ }{\ }{\ }
   {\ }{\ }{\ }{\ }  \lac \GG^a , \BB_{\S} \rac = \HH^a \ ,
   {\ }{\ }{\ }{\ }  \lc \HH^a , \GG^b \rc = - f^{abc} \GG^c \ , \es
   \lc \HH^a , \BB_{\S} \rc = 0 \ ,
   {\ }{\ }{\ }{\ }  \lc \BB_{\S} , \VV \rc = - \UU \ ,
   {\ }{\ }{\ }{\ }  \lc \HH^a , \HH^b \rc = -f^{abc} \HH^c  \ , \es
\eqacn{algebrbbvv}
which, taken together with \equ{offsusy}, show that the operators
$(\BB_{\S}, \WW_\m, \PP_\m, \GG^a, \HH^a, \VV, \UU)$ form a closed
algebra.

Let us conclude this section by displaying in Table~\ref{dim1-gh1}
the dimension and the ghost number of all the fields and sources.
\begin{table}[hbt]
\centering
\begin{tabular}{|c|c|c|c|c|c|c|c|c|c|c|c|c|}
\hline &$A_\m$&$b$&$\bar{c}$&$c$&$D$&$\hat{D}$&$K$&$\bar{K}$
&$\la^\m$&${\bar \lambda}^\m$&$\g^\m$&$\s$\\
\hline  dim &1&2&2&0&1&1&1&1&-1&-1&2&3\\
\hline  ghost\# &0&0&-1&1&0&1&0&-1&1&2&-1&-2\\ \hline
\end{tabular}
\caption[t1]{Dimensions and ghost numbers}
\label{dim1-gh1}
\end{table}
%
%
\section{Renormalization and finiteness}
In order to study the ultraviolet behaviour of the Chern-Simons theory
in the interpolating gauge let us begin by showing that the
Slavnov and the supersymmetric Ward identities
\equ{slatay}, \equ{susy} as well as the symmetries (iii),(iv) and (v) are not
anomalous.
Following \cite{susy}, this is easily done by collecting all the
operators into a unique nilpotent operator by means of
additional global ghost parameters
$(\th,\e,\a,\b,x,y)$ whose dimensions and ghost number are given
in Table~\ref{dim2-gh2}.
\begin{table}[hbt]
\centering
\begin{tabular}{|c|c|c|c|c|c|c|}
\hline &$\th^\la$ &$\e^\la$ & $x^a $& $y^a$ &$ \a$ &$ \b$ \\
\hline dim & -1& -1& 0 &0& -1& -1\\
\hline ghost\# & 2& 1& 2& 1& 2& 3\\
\hline
\end{tabular}
\caption[t2]{Dimensions and ghost numbers of the global ghosts}
\label{dim2-gh2}
\end{table}

Let us introduce then the nilpotent operator $\QQ$ defined as
\eqac
\QQ\= \BB_{\S}+\th^\la \WW_\la + \e^\la \PP_\la+x^a \GG^a+
y^a \HH^a + \a \UU +\b \VV \es
&&- (x^a-\frac{1}{2}f^{abc} y^b y^c )\dpad{}{y^a}-f^{abc}x^b y^c \dpad{}{x^a}
- \th^\la \dpad{}{\e^\la} -\b\dpad{}{\a} \ ,\es
\eqacn{Qtrafo}
\eq
  \QQ \QQ = 0  \ .
\eqn{Q-nilp}
It is apparent now that the operator $\QQ$ collects all the
symmetries of the action $\S$. Moreover, as shown in \cite{susy}, the
absence of anomalies for the operator $\QQ$  is equivalent to
the absence of anomalies for each single operator entering the
expression \equ{Qtrafo}.

Let us consider then the consistency condition
\eq
   \QQ \AA = 0 \ ,
\eqn{QQ-consistency}
where the possible anomaly $\AA$ is an integrated local polynomial of
dimension three and ghost number one. To study the cohomology of
$\QQ$ we introduce the filtering operator $\NN$
\eqac
\NN\=  y^a\dpad{}{y^a}+x^a\dpad{}{x^a}+\th^\la\dpad{}{\th^\la}+
\e^\la\dpad{}{\e^\la } +\a\dpad{}{\a} +\b\dpad{}{\b}\es
\eqacn{qn}
according to which the operator \equ{Qtrafo} decomposes as
\eq
\QQ = \QQ^{(0)} + \QQ^{(R)}\,,
\eqn{decomposition}
with
\eq\ba{lcl}
\lc \QQ^{(0)} , \NN\rc = 0 \ , \\
\QQ^{(0)} = \BB_\S -   x^a\dpad{}{y^a}  - \th^\la\dpad{}{\e^\la} -
                       \b \dpad{}{\a} \ .
\ea\eqn{dnull}
Since the cohomology of $\QQ$ is isomorphic to a subspace of the cohomology
of $\QQ^{(0)}$ we focus on this last operator.
{}From \equ{dnull} one immediately sees that the ghosts
$(x,y,\theta,\e,\beta,\alpha)$ are grouped in BRST-doublets which are known
to yield a vanishing cohomology. Therefore the characterization of the
cohomology of the operator $\QQ^{(0)}$ reduces to that of the
Slavnov-Taylor operator $\BB_\S$.
However, due to fact that in three dimensions there are no gauge anomalies,
it follows that $\QQ^{(0)}$ has vanishing cohomology in the
sector of field polynomials of dimension three and ghost number one,
implying then that the consistency condition \equ{QQ-consistency} has
only trivial solutions, i.e.
\eq
    \AA  =  \QQ {\hat \AA}
\eqn{trivial}
We have proven thus the absence of anomalies for all the operators
entering in \equ{Qtrafo}.

Let us proceed now to the analysis of the invariant local
counterterms $\G^{\rm count}$ corresponding to possible
coupling constant and field amplitudes renormalization.
The gauge conditions \equ{gaugecon} and the anti-ghost equations
\equ{antighost} imply that $\G^{\rm count}$ does not depend on
$(b, D, {\hat D}, K)$ and that the fields $\bar K$, $\gamma$ and
$\bar c$ enter only in the combination
\eq
 \OM^\m = \g^\m- n^\m \bar{c} -\pa^\m {\bar K} \ .
\eqn{combination}
Taking into account also the Slavnov identity \equ{slatay} and the
ghost equation \equ{ghost}, it
follows that
the most general local invariant counterterm can be parametrized as
\eq\ba{rl}
\G^{\rm count} = &\!\! \tr \dint\,d^3x\,\Lp -\frac{a}{2}\e^{\m\n\r}\lp
A_\m\pa_\n A_\r +\frac{2}{3}A_\m A_\n A_\r\rp\Rp  \\
 &\!\!
 +{\ }\BB_{\S} \tr\dint\,d^3x\, \Lp \a^\m_{\  \n} A_\m \OM^\n
    +\b_{\m\n\r}\OM^\m \OM^\n
   \la^\r+ \tau^\m_{\  \n}A_\m\s \la^\n  \\
 &\!\!
  {\ }{\ }{\ }{\ }{\ }{\ }{\ }{\ }{\ }{\ }{\ }{\ }{\ }{\ }{\ }{\ }
  + \d \e_{\m\n\r}\s^2 \la^\m\la^\n\la^\r
   +\om_{\m\n\r}\OM^\m \s \la^\n\la^\r \Rp \ ,
\ea\eqn{count}
where
$(a,\alpha^\mu_\nu,\b_{\mu\nu\rho},\tau^\mu_\nu,\delta,\omega_{\mu\nu\rho})$
are arbitrary quantities which depend on the vectors $(n,n^*)$.

Applying now the supersymmetric Ward identity \equ{susy} we get the
homogeneous condition
\eq
  \WW_\mu \G^{\rm count} = 0  \ ,
\eqn{susy-condition}
from which it follows
\eqac
\a^\m_{\  \n}\= \tau^\m_{\  \n}\; = -\d^\m_{\   \n}\lp \frac{a}{3}
        +2 \d\rp\es
\om_{\m\n\r}\= 3\d \e_{\m\n\r}\es
2\b_{\m\n\r}\= \lp -\frac{a}{3} +2 \d\rp\, \e_{\m\n\r}
\eqacn{param}
so that $\G^{\rm count}$ takes the following restricted form
\eq\ba{rl}
\G^{\rm count} = \tr\dint &\!\!
         \,d^3x\, {\bar a}\Lp \half \e^{\m\n\r} A_\m \pa_\n A_\r
+\OM^\m\pa_\m c - \pa_\m c\s \la^\m + A_\m \s {\bar \la}^\m -
D_\r\OM^\r A_\m\la^\m \\
 &\!\!
  + \lp \pa_\m A_\r-\pa_\r A_\m +\lc A_\m,A_\r \rc\rp \OM^\r \la^\r+
    \half \e_{\m\n\r}\OM^\m \OM^\n {\bar \la}^\r\Rp \\
 &\!\!
  + \d \Lp 3\e^{\m\n\r} A_\m \pa_\n A_\r +
2 \e^{\m\n\r} A_\m A_\n A_\r +3\e_{\m\n\r}\s^2\la^\m\la^\n\la^\r \\
 &\!\!
    +6  \lp \pa_\m A_\r-
\pa_\r A_\m +\lc A_\m,A_\r \rc\rp\OM^\r \la^\m  \\
 &\!\!
  +3 \e_{\m\n\r}\OM^\m\OM^\n {\bar \la}^\r-
  3 \lp \pa_\m A_\r-
  \pa_\r A_\m +\lc A_\m,A_\r \rc\rp
  \s\la^\m\la^\r  \\
 &\!\!
  +6\e_{\m\n\r}\OM^\m\s\la^\n {\bar \la}^\r-
3\e_{\m\n\r}D_\e\OM^\e\OM^\m\la^\n\la^\r-
2\e_{\m\n\r}D_\la\OM^\la \s\la^\m\la^\n\la^\r \Rp
\ea\eqn{count2}
whith $\bar a$ defined as $({\bar a}\equiv -\frac{a}{3}-2\d)$.

It remains now to impose the two transversality conditions \equ{transcon}.
It is very easy to see that these conditions enforce the vanishing of all
the terms of \equ{count2} containing the external sources
$\la^\m$ or ${\bar \la}^\m$. Therefore
\eq
  a = 0  \ , \qquad \d= 0 \ ,
\eqn{a-d-condition}
and
\eq
   \G^{\rm count} = 0 \ ,
\eqn{van-countert}
meaning that there is no possible local counterterm compatible with
the symmetries and constraints of the classical action.
This proves the perturbative finiteness of the Chern-Simons theory
in the interpolating gauge \equ{CSfix}.

%
%

\end{document}